\documentclass[twocolumngrid,prd,superscriptsize,reprint]{revtex4-1}
\usepackage[]{graphicx}
\usepackage{dcolumn}
\usepackage{bm}

\usepackage{amsmath,amssymb,amsfonts}
\usepackage{fancyhdr}
\usepackage{lipsum} 
\usepackage{subcaption}
\usepackage{multirow}

 \usepackage[colorlinks]{hyperref}
 \hypersetup{
     colorlinks = false,
   }

\newcommand{\comment}[1]{}

\usepackage{color}
\usepackage{babel}
\newcommand{\bea}{\begin{eqnarray}}
\newcommand{\eea}{\end{eqnarray}}

\newcommand{\be}{\begin{equation}}
\newcommand{\ee}{\end{equation}}

\begin{document}

\title[]{Moduli space with a boundary}

\author{C. Adam}
\email[]{adam@fpaxp1.usc.es}
\affiliation{Departamento de F\'isica de Part\'iculas, Universidad de
Santiago de Compostela and \\
Instituto Galego de F\'isica de Altas Enerxias (IGFAE), E-15782
Santiago de Compostela, Spain}

\author{K. Oles}
\email[]{katarzyna.slawinska@uj.edu.pl}
\author{T. Romanczukiewicz}
\email[]{tomasz.romanczukiewicz@uj.edu.pl}
\author{A. Wereszczynski}
\email[]{andrzej.wereszczynski@uj.edu.pl}
\affiliation{Institute of Theoretical Physics, Jagiellonian University,
Lojasiewicza 11, Krak\'{o}w, Poland}

\begin{abstract}
We find that for various solitonic processes the corresponding canonical moduli space can have a boundary which is accessible in a finite time evolution. We show that such a boundary is not a failure of the moduli space approach but has a physical meaning. In our example, it corresponds to the complete annihilation of a colliding kink and antikink after a finite time.  We further find that, close to the boundary, the solutions have an approximate self-similar form. 
\end{abstract}
\maketitle

\section{Motivation}

The {\it moduli space} (MS) is an important concept in the dynamics of topological solitons, where it allows to approximate the full evolution provided by the original Lagrangian $L[\phi]$ by the simpler dynamics following from a collective coordinate model (CCM) $L[a^i]$ \cite{NM-1}. In this framework, the infinite-dimensional space of all field configurations is replaced by a restricted set of configurations which are spanned by a finite number of parameters $a^i$, $i=1..N$, called collective coordinates or {\it moduli}. They give rise to the moduli space, which is a Riemannian manifold $\mathcal{M}$ with a metric $g$ inherited from the kinetic part of the original model, see \cite{MS} for a review. In the best understood Bogomol'nyi–Prasad–Sommerfield (BPS) case \cite{Bogomolny}, the moduli space is built using energetically equivalent BPS solutions, \cite{NM-1, BPS-moduli-1, BPS-moduli-2}. This MS is usually referred to as the {\it canonical moduli space}. To correctly describe the dynamics of solitons, it is often necessary to go beyond the set of energetically equivalent solutions and include configurations which differ in energy. The inclusion of massive internal modes, e.g.,  results in the {\it vibrational moduli space}. 

Ideally, such a space should be globally well defined without any singularities or boundaries which could be accessible in a finite time evolution. Indeed, their existence would lead to a breakdown of the CCM dynamics after a finite time. It was, therefore, assumed that the appearance of a singularity or a boundary is an unacceptable feature of a MS \cite{MM}. 

However, it has been recently pointed out that an essential singularity of a MS can have an important physical meaning. In fact, it can occur in a vibrational moduli space where it signals the appearance of a {\it spectral wall} \cite{SW, SW-2field}. Strictly speaking, a spectral wall is an obstacle in the motion of a topological soliton due to the transition of a normal mode (vibrational mode) to the continuum spectrum. Such an obstacle results in the formation of an arbitrarily long living unstable stationary solution representing, e.g., a pair of a kink and an antikink frozen at a certain mutual distance and performing small oscillation around their positions. The separation between the solitons is fixed by the point where the mode hits the mass threshold. After the discovery of spectral walls in BPS-impurity systems \cite{SW}, they have also been found in two-scalar field theories \cite{SW-2field} and in kink scatterings in models with different mass vacua as, e.g., in antikink-kink collisions in the $\phi^6$ model, \cite{SW-phi6}. In any case, from the point of view of the vibrational moduli space, at this point a singularity is developed and some of the metric functions diverge. Of course, this singularity is related to the fact that the mode ceases to be a normalizable mode at this point. 

It should be underlined that the application of the moduli space concept goes beyond an approximation of the full theory in terms of a collective coordinate model. In fact, it is a very useful tool which allows for the explanation of various phenomena observed in solitonic collisions like, for example, the existence of a fractal structure in the final state formation in multi-kink collisions due to the resonant energy transfer \cite{sug, CFW}, or the $90^
\circ$ scattering of solitons \cite{BPS-moduli-1, BPS-moduli-2, vortex-1, AM}. The MS concept is also of utmost importance for quantization, where the metric on the moduli space defines a Laplace-Beltrami operator which is the central ingredient of the semiclassical approach to quantum solitons \cite{GM, Hal-1, Hal-2, Gud}. 

In the present paper, we show that also the existence of an accessible boundary should not be necessarily treated as a failure of the moduli space approach. On the contrary, such a boundary may again have a physical importance denoting, for example, the complete annihilation of a kink-antikink pair after a finite time. In our example, we use a particular BPS-impurity model. However, as we discuss later, the same property also exists in more general models like, e.g., multi-scalar field theories.

\section{Collective coordinates and moduli space}

Here we briefly describe the collective coordinate/moduli space framework. For simplicity we consider a single scalar field field theory in (1+1) dimensions defined by the following Lagrangian
\be
L[\phi]=\int_{-\infty}^\infty \left( \frac{1}{2} \phi_t^2 - \frac{1}{2} \phi_x^2 - U(\phi) \right) dx, \label{Lag-scalar}
\ee
where $\phi_t, \phi_x$ are time and space derivatives and $U(\phi)$ is a field theoretical potential, which is assumed to have at least two vacua to guarantee the existence of topological solitons. 

In the CCM approach, the infinite dimensional set of all possible field configurations is reduced to an $N$ dimensional set $\mathcal{M}[{\bf a}]=\{ \Phi(x; a^i), i=1..N \}$, which should capture the most important properties of the particular solitonic process we want to describe. The construction of this set is the main part in the framework and, at the same time, it is also the most difficult step. The reason is that, for an arbitrary process, there is no canonical way to build it. A natural guess is to include the lowest energy excitations, i.e., {\it (i)} all energetically equivalent static solutions where the initial and final states are necessary included and {\it (ii)} the lightest massive excitations, e.g., normal massive modes carried by single soliton states as well as by pertinent multi-soliton configurations. Here other massive excitations like Derrick modes or quasi-normal modes may also be relevant. A moduli space based only on configurations of the first type is called the canonical moduli space and has been proved to work quite well for many BPS processes. The extension to configurations of the second type provides a vibrational moduli space.

Once the restricted set of configuration is chosen, we promote all collective coordinates to time dependent variables ${\bf a}(t)$. Then, we insert the resulting configurations into the original Lagrangian and perform the spatial integration. This leads to an effective CCM 
\be
L[{\bf a}]= \frac1{2} g_{ij} ({\bf a})\; \dot{a}^i \dot{a}^j  - V ({\bf a}),
\ee
where $g_{ij}$ is the moduli space metric
\be
g_{ij} ({\bf a}) = \int_{-\infty}^\infty  \frac{\partial \Phi (x; {\bf a})}{\partial a^i}  \frac{\partial \Phi (x; {\bf a})}{\partial a^j}   dx
\ee
while $V$ is the effective potential 
\be
V ({\bf a}) = \int_{-\infty}^\infty \left( \frac{1}{2} \left( \frac{\partial \Phi (x; {\bf a})}{\partial x} \right)^2+ U( \Phi (x; {\bf a}) ) \right) dx .
\ee
The simplest example of this construction corresponds to the one-dimensional moduli space built from the static energetically equivalent kink (or antikink) solutions at all possible locations $\mathcal{M}= \{ \Phi_{K(AK)} (x-a); a \in \mathbb{R} \}.$ Here $\Phi_{K(AK)} (x-a)$ solves the usual Bogomolny'i equation 
\be
\phi_x = \pm  \sqrt{2U}. \label{Bog-eq}
\ee
Now, the moduli space is just the real line equipped with the constant metric and constant effective potential $g_{aa}=V=M$, where $M = \int_{-\infty}^\infty  U( \Phi )  dx$ is the mass of the soliton. A solution of such a CCM is a non-relativistic 
 kink of constant velocity. Indeed, $a(t)= vt +a_0$. In order to get a relativistic motion of the kink one needs to extend the moduli space to a two-dimensional case, $\mathcal{M} = \{ \Phi_{K(AK)} (b(x-a)); a \in \mathbb{R}, b \in \mathbb{R}_+ \}$, where $b$ is the scale deformation parameter \cite{Rise, Caputo, RMS}. Obviously, the energy increases for $b\neq 1$. Then, the metric has two non-trivial components $g_{aa}, g_{bb}$ which are simple functions of the parameter $b$. The solution of the corresponding CCM is $\dot{a}=v, b=1/\sqrt{1-v^2}$, which leads to a relativistic boosted kink. 

This can be further developed to describe more interesting multi-kink collisions. In fact, it has been very recently proven that a CCM reproduces the qualitative and even quantitative properties of the kink-antikink scattering in the $\phi^4$ model. In this case, one needs to take into account a vibrational mode (shape mode) \cite{MORW} or Derrick modes \cite{RMS}. Importantly, the corresponding moduli spaces are smooth manifolds without any singularities or boundaries. Possible points where metric functions diverge or vanish were proven to be apparent singularities which can be removed by a suitable choice of coordinates  \cite{MORW, MORW-2}. Other unremovable singularities as well as possible boundaries of $\mathcal{M}$ are not attainable by finite energy configurations, because the effective potential diverges at these points. 

It was, therefore, a surprise that for some processes the vibrational moduli space may have singularities which are not merely an artifact of badly chosen collective coordinates but, on the contrary, do have a real impact on the solitonic dynamics, giving rise, e.g., to the spectral wall phenomenon \cite{SW}. 

In the subsequent analysis we show that for some solitonic processes the moduli space can have a boundary which is attained in a finite time. Importantly, we claim that this is not a disqualifying feature of the MS but, instead, can have a clear physical meaning. As an example, we use the particularly simple case of BPS-impurity models but, as we discuss later on, the same phenomenon can be expected in much more general systems.

\section{Moduli space with accessible boundary}

We begin with the sine-Gordon squared model
\be
L[\phi]=\int_{-\infty}^\infty \left( \frac{1}{2} \phi_t^2 - \frac{1}{2} \phi_x^2 - \frac{1}{2}(1-\cos \phi)^2 \right) dx.
\ee
Hence the potential is $U_{sG^2}=(1/2)(1-\cos \phi)^2$. 
The model has a symmetry $\phi \to \phi +2\pi$ and therefore the fundamental domain of the scalar field is $[0,2\pi]$ with the vacua $\phi=0$ and $\phi=2\pi$. It is also convenient to introduce a prepotential $W$ such that $U=(1/2)W^2$. Thus, here 
\be
W_{sG^2}=1-\cos \phi. 
\ee
There is a kink and antikink interpolating between the vacua
\be
\Phi_{K(AK)} = \mp \, \left( \pi + 2\arctan \left( x -a\right) \right) .
\ee
They verify the corresponding Bogomol'nyi equation
\be
\phi_x = \pm (1-\cos \phi).
\ee
We remark that these solitons are examples of kinks with fat (or slowly decaying) tails, see for example \cite{fat-1, fat-2, fat-3}. Indeed, the behaviour of the kink close to the vacua is $\Phi_K \approx 2\pi - \frac{2}{x}$  for $x \to \infty$ and $\Phi_K \approx - \frac{2}{x}$ for $x \to -\infty$. An important consequence of that is that, in contrast to exponentially localized solitons, a moduli space for antikink-kink collisions cannot be built from the naive sum of the single soliton states. Such a naive superposition is also a very bad initial configuration for a numerical scattering, because it implies the addition of a huge amount of energy stored between the colliding solitons \cite{Gani}. 

Now, let us introduce a BPS-impurity version of a general single scalar field model (\ref{Lag-scalar}) and apply it to the squared sG model. This means that we couple it with a background (non-dynamical) field $\sigma$ in the following way \cite{solvable}
\be
{L}[\phi,\sigma]=\int_{-\infty}^\infty \left( \frac{1}{2}\phi_t^2-\frac{1}{2}\left( \phi_x - \sigma W\right)^2 -\phi_xW \right)dx. \label{BPS-imp}
\ee
The crucial feature of this class of impurity models is that it still possesses {\it one} Bogomol'nyi equation. Indeed, we find that
\bea
E&=&\int_{-\infty}^\infty \left(  \frac{1}{2}\phi_t^2+\frac{1}{2}\left( \phi_x - \sigma W\right)^2 +\phi_xW\right) dx \nonumber \\
&\geq & \int_{-\infty}^\infty \phi_xW dx =  \int_{\phi(-\infty)}^{\phi(\infty)} Wd\phi 
\eea
The inequality is saturated if and only if the following impurity modified Bogomol'nyi equations are satisfied
\be
\phi_t=0, \;\;\;  \phi_x - \sigma W =0 \label{Bog-imp}
\ee
It is easy to verify that solutions of the Bogomol'nyi equations obey the second order equation of motion
\be
-\phi_{tt}+\phi_{xx} -\sigma_x W-\sigma^2 WW_\phi=0 \label{eom}
\ee
and, therefore, are static solutions of the impurity deformed model minimizing the energy in a corresponding topological sector.
Observe that for $\sigma=1$ we obtain the original theory without impurity. 

\begin{figure}
 \includegraphics[width=1.00\columnwidth]{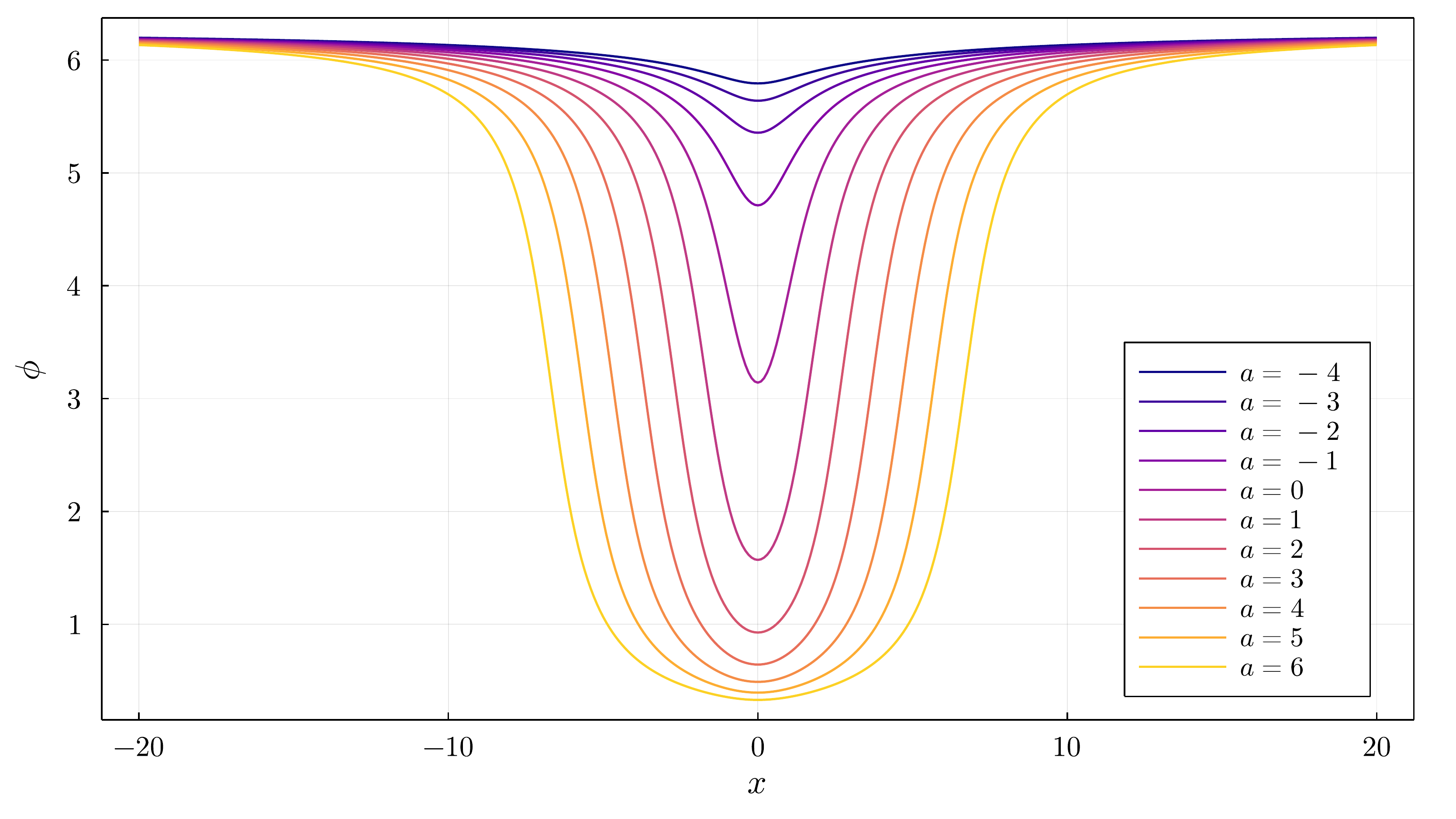}
 \caption{BPS solutions of the BPS-impurity model with $W_{sG^2}$. }\label{plot-moduli-boundary}
 \end{figure}
 \begin{figure}
 \includegraphics[width=1.00\columnwidth]{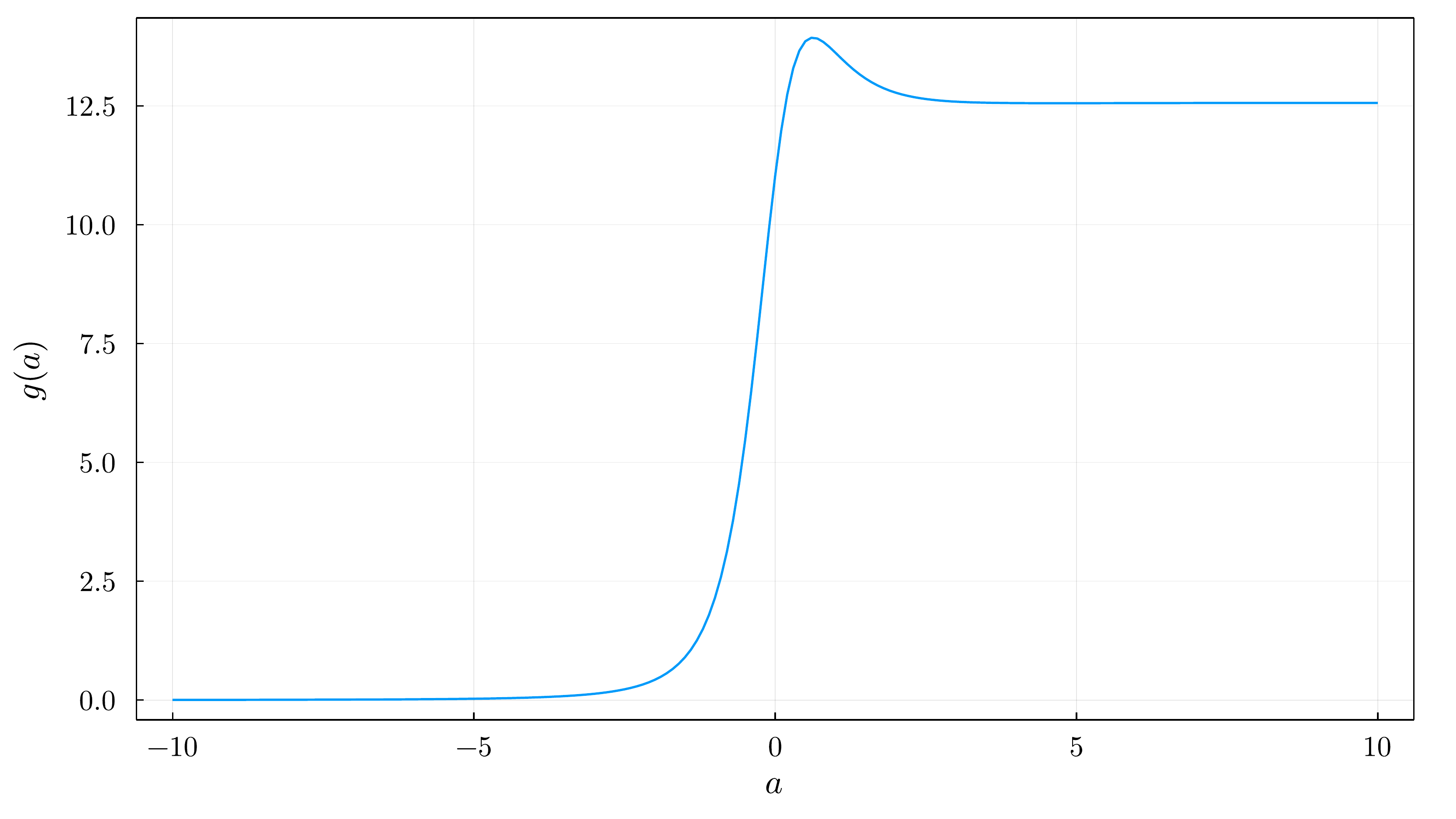}
 \caption{Moduli space metric $g(a)$.}\label{plot-moduli}
 \end{figure}

For our purposes we choose the impurity in the form
\be
\sigma = \tanh(x).
\ee
Then, the Bogomol'nyi equation can support a pair of antikink and kink (or kink and antikink) as a BPS solution. Indeed, for $x \to \mp \infty$, the impurity approaches $\mp 1$, and we get the antikink and kink Bogomol'nyi equation, respectively. Strictly speaking we find a one-parameter family of BPS (energetically equivalent) solutions. This allows us to study kink collisions in the BPS limit where there is no static force between the solitons and where the canonical moduli space works very well. 

For $W=W_{sG^2}$ the Bogomolny'i equation reads
\be
\phi_x = 2 \tanh(x)  \sin^2 \frac{\phi}{2} \label{Bog-imp-sG2}
\ee
leading to the following energetically equivalent BPS solutions
\be
\Phi_{AKK}(x;a)=
\pi + 2\arctan \left( \ln (\cosh(x)) -a\right) 
\ee
Note that the $2\pi$ symmetry remains unchanged. We present the field profiles in Fig. \ref{plot-moduli-boundary}. We start with $a \to \infty$ and find a pair of (infinitely) separated antikink and kink. As $a$ tends to 0 and then further decreases to negative values, the solitons approach each other and, finally, for $a \to -\infty$ they form the $\phi=2\pi$ vacuum. Thus the moduli space is just a real line $\mathcal{M} = \{ \Phi_{AKK}(x;a); a \in \mathbb{R} \}$ with the metric
\be
g(a)=4 \int_{-\infty}^\infty \frac{dx}{(1+(\ln(\cosh(x)) -a)^2)^2}
\ee
We plot the metric in Fig \ref{plot-moduli}. We see that it  interpolates between $g(a=\infty)=4\pi$ and $g(a=-\infty)=0$. 

The crucial observation is that the boundary at $a=-\infty$ has a finite distance from any other point on the moduli space, $a\in \mathbb{R}$ and therefore it can be attained in a finite time. 
 
 \begin{figure}
 \includegraphics[width=1.10\columnwidth]{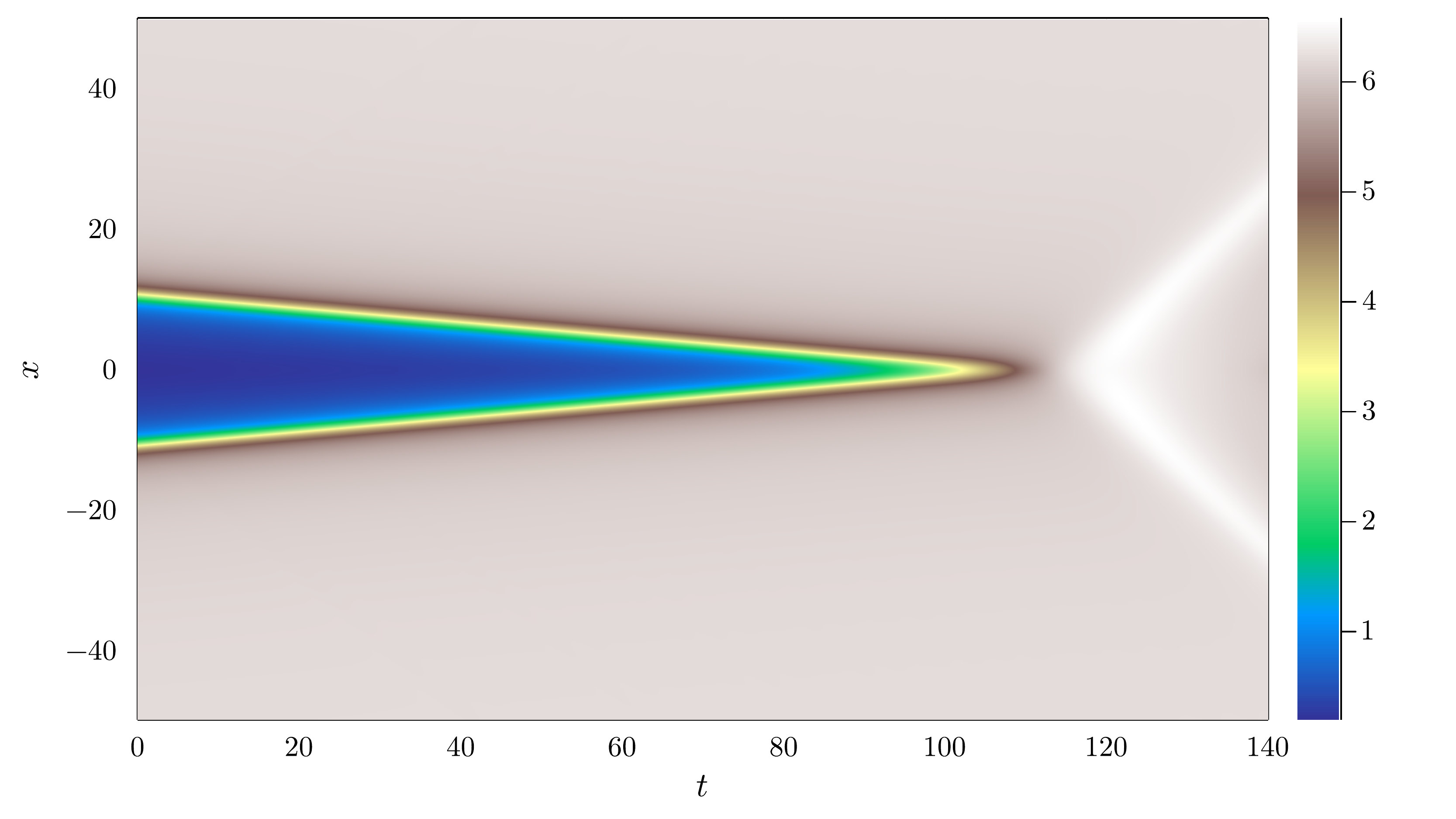}
 \caption{Antikink-kink annihilation in the BPS-impurity model with $W_{sG^2}$.}\label{plot-imp-BPS}
 \includegraphics[width=1.00\columnwidth]{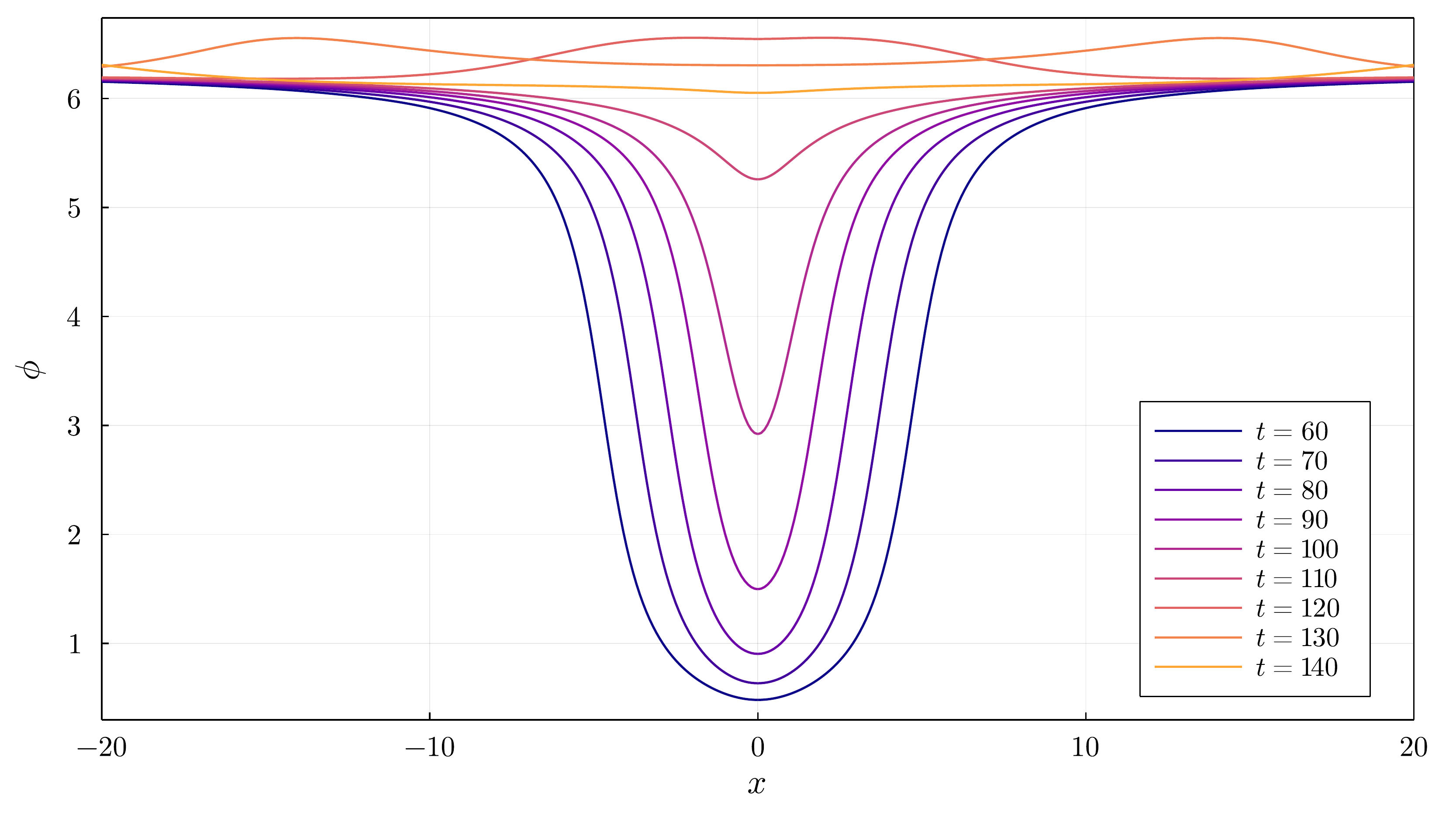}
 \caption{Profiles of the field in the antikink-kink scattering in the BPS-impurity model with $W_{sG^2}$.}\label{plot-imp-BPS-2}
 \end{figure}

To see this, we consider a natural $L^2$ metric on the space of real functions. Namely, the squared distance between two functions $\phi_1, \phi_2$ is 
\be
s^2=\int_{-\infty}^\infty \left( \phi_1(x)-\phi_2(x)\right)^2 dx
\ee
In our case, the distance between any member of the moduli space, $\Phi_{AKK}(x;a)$ and the boundary at $a=-\infty$, i.e., $\Phi_{AKK}(x;-\infty) = 2\pi$ is a finite number. Indeed, 
\be
\int_{-\infty}^\infty \left( \pi - 2\arctan \left( \ln (\cosh(x)) -a\right)   \right)^2 dx
\ee
is a convergent integral since at $x \to \pm \infty$ the integrand function $(\pi - 2\arctan \left( \ln (\cosh(x)) -a\right) )^2\approx \frac{4}{x^2}$. 

Equivalently we can analyze the metric close to the boundary $a=-\infty$. It behaves as $g(a) \approx \frac{8}{3} a^{-3}$. Thus the effective CCM reads
\be
L [a] \approx \frac{4\dot{a}^2}{3a^{3}}, \;\;\; a\to -\infty.
\ee
This can be brought to a trivial form of a flat metric by the following change of variable, $a=-b^{-2}$. Indeed, $L[b] \sim \dot{b}^2$ and the solutions are $b=v_bt +b_0$. Thus finally, $a = - (v_b t+b_0)^{-2}$, and for a finite $v_b$ and $b_0$ the modulus $a$ diverges at a finite time.
\begin{figure}
 \includegraphics[width=1.10\columnwidth]{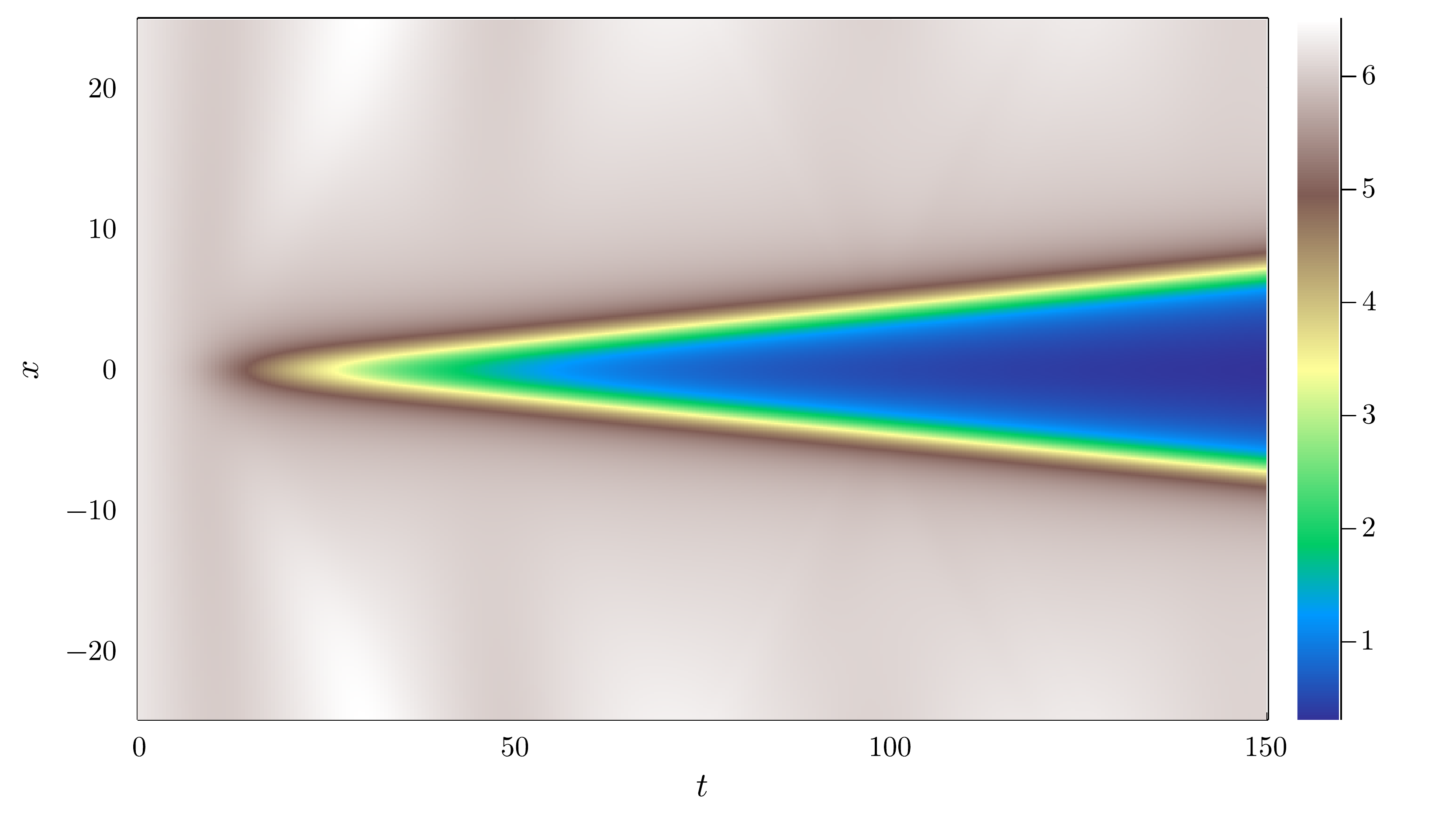}
 \caption{Antikink-kink creation in the BPS-impurity model with $W_{sG^2}$.}\label{plot-imp-BPS-cr}
 \includegraphics[width=1.00\columnwidth]{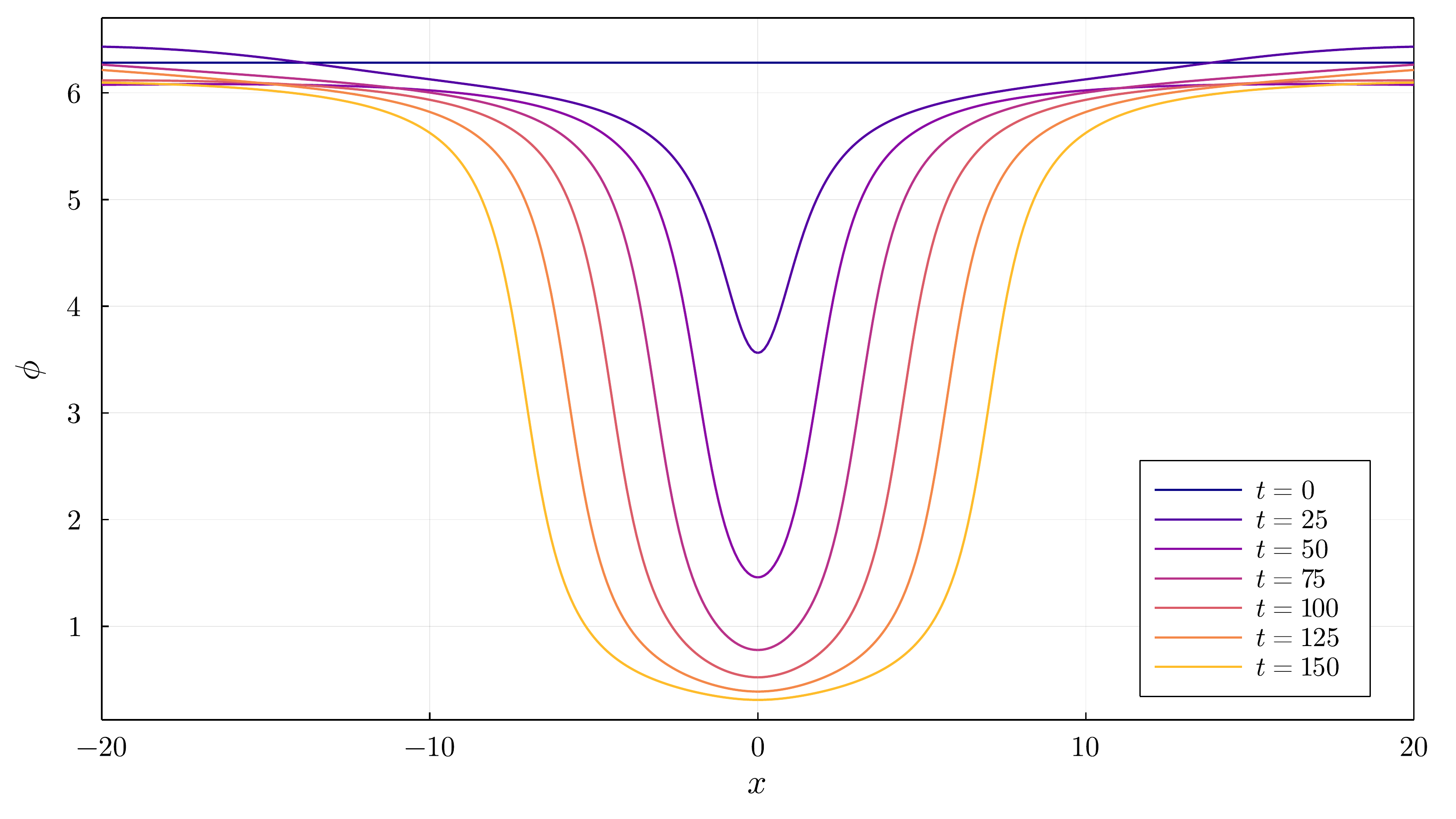}
 \caption{Profiles of the field in the antikink-kink creation in the BPS-impurity model with $W_{sG^2}$.}\label{plot-imp-BPS-2-cr}
 \end{figure}
 
In conclusion, the moduli space has an attainable boundary. We remark that there are no other solutions of the Bogomolny'i equation (\ref{Bog-imp-sG2}) which could be used to extend the MS. Therefore whatever configurations we use they will necessarily have higher energy. 

At a first glance, this result could be treated as a serious drawback, not only of the presented MS but even the whole CCM approach, as it proves that there are models for which the canonical construction of the MS leads to a CCM which breaks down in a finite time. However, a closer look into the actual antikink-kink collision in the full BPS-impurity model (\ref{eom}) sheds a new light on our results. 

In Fig. \ref{plot-imp-BPS} we show the full time evolution of the following initial configuration, representing a collision of the antikink and kink, 
\bea
\phi(x,0) &=& \pi + 2\arctan \left( \ln (\cosh(x)) -a_0\right) \nonumber \\
\partial_t \phi(x,0) &=& \frac{2v}{1+\left( \ln (\cosh(x)) -a_0\right)^2} 
\eea
where $a_0$ is related to the initial separation of the solitons and $v$ is their initial velocity. We take $a_0=10$ and $v=0.1$.  In Fig. \ref{plot-imp-BPS-2} we also show profiles of the field at several $t$.  In our numerical analysis we clearly see that the solitons approach each other and then, after a finite time, they completely annihilate to the vacuum. This occurs exactly as predicted in the moduli space framework. As the antikink and kink disappear, the excess of energy (e.g., stored in the kinetic energy of the colliding solitons) is radiated in two lumps emitted with the speed of light. This radiative process, obviously, is not covered by the CCM constructed above, based on the simplest one-dimensional moduli space. 

Of course, the boundary of the moduli space can be related to a reverse process, that is, an antikink-kink creation. In Fig. \ref{plot-imp-BPS-cr} we show how such a pair is produced from an initial lump of energy. Its evolution very precisely follows the geodesic dynamics of the moduli space. The initial configuration is $\phi(x,0)=2\pi$ while $\partial_t \phi(x,0)=\Phi_{AKK}(x;a= -40)-2\pi$, which corresponds to an initial lump of the energy density located at the origin. In Fig. \ref{plot-imp-BPS-2-cr} we show the snapshots of the field at some $t$. 

\begin{figure}
 \includegraphics[width=1.10\columnwidth]{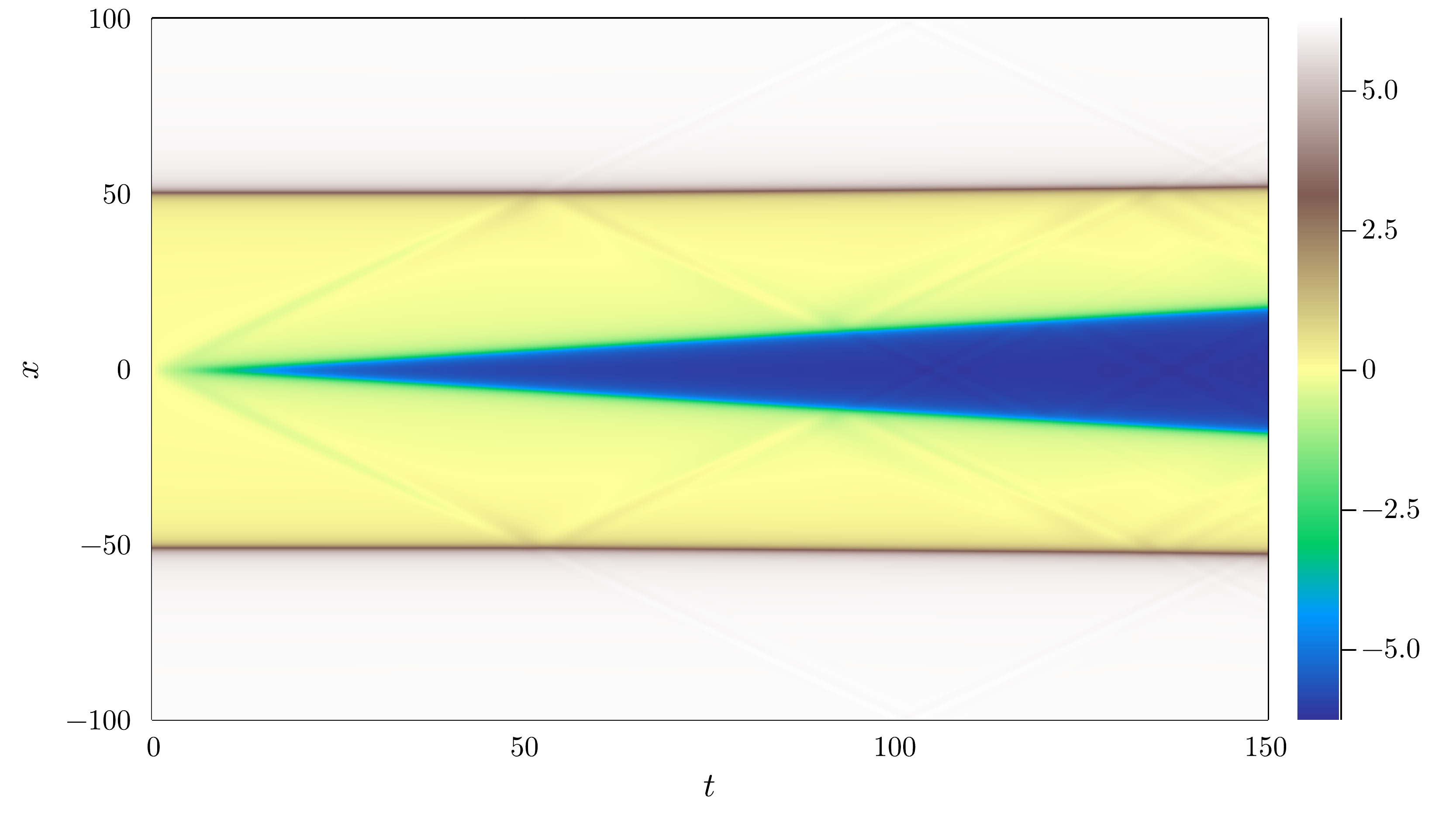}
 \caption{Antikink-kink creation between a separated antikink-kink pair in the BPS-impurity model with $W_{sG^2}$.}\label{plot-imp-BPS-cr-2}
 \includegraphics[width=1.00\columnwidth]{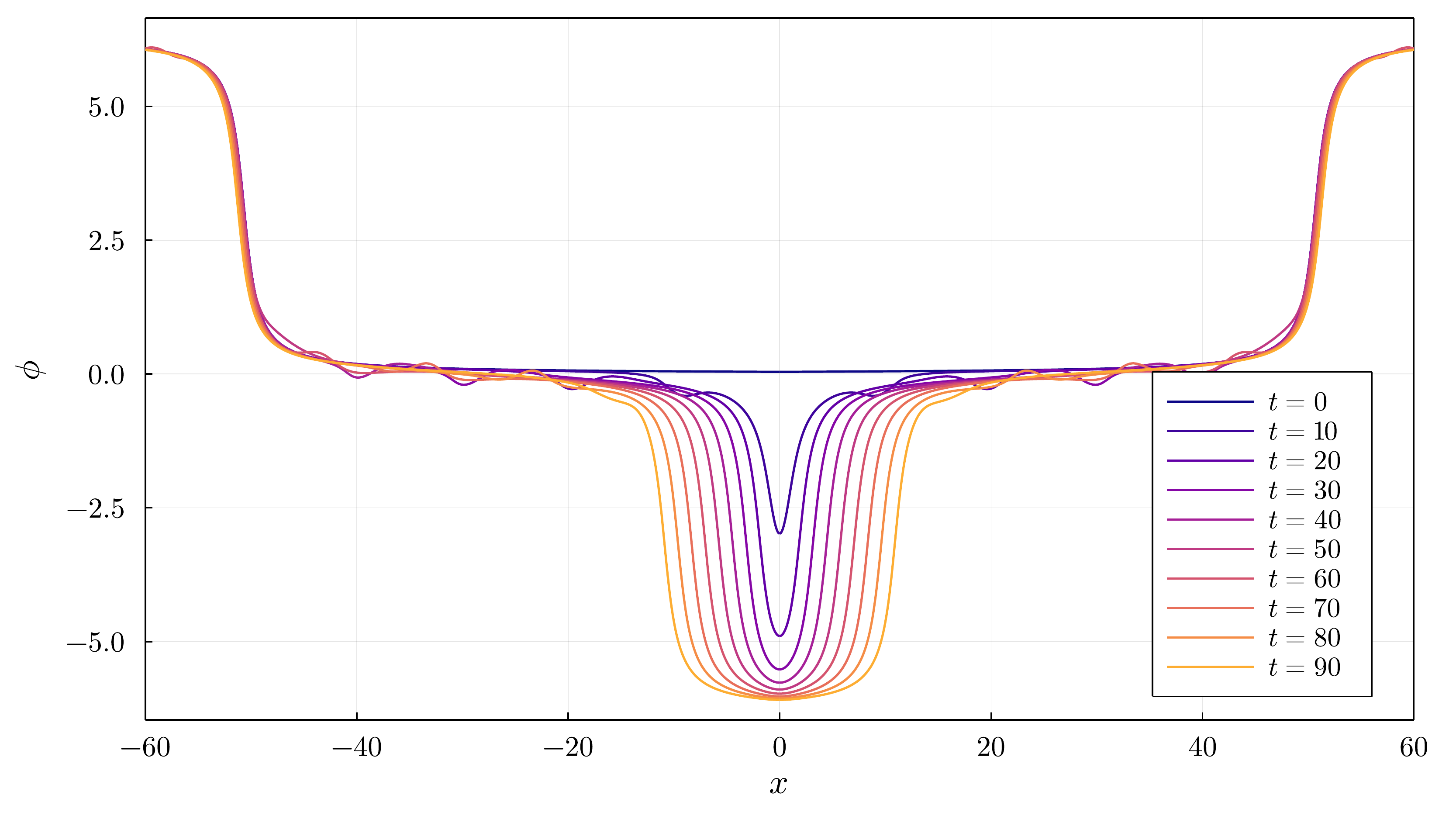}
 \caption{Profiles of the field in the antikink-kink creation between a separated antikink-kink pair in the BPS-impurity model with $W_{sG^2}$.}\label{plot-imp-BPS-2-cr-2}
 \end{figure}

This leads to the following property. Whenever the field is close to the vacuum there is a tendency to create an antikink-kink pair from a small energy surplus. This is presented in Fig. \ref{plot-imp-BPS-cr-2} and Fig. \ref{plot-imp-BPS-2-cr-2}. 

It is important to remark that not all initial conditions immediately lead to the creation of an antikink-kink pair. On the contrary, one has to prepare a specific lump of energy to observe such a creation process. For example, if we start with an exponentially localized initial profile $\phi(x,0)=-C/\cosh(x), \phi_t(x,0)=0$, where $C>0$, then it mainly decays into two energy lumps, wave packets, which escape with the speed of light to spatial infinity. This is the main energy transfer channel. Around the origin, there is still a small fraction of energy stored in the form of a wave with a very small wave vector, whose oscillations finally create an antikink-kink pair. This scenario repeats for other initial configuration e.g., of a gaussian form. This will be explained in the next section. 

To summarize, we have shown that there are solitonic processes which admit a collective coordinate description with a moduli space whose boundary is approached in a finite time. Furthermore, this boundary has a physical meaning, indicating a complete annihilation of the solitons to the vacuum in a finite time. 

\section{Self-similarity }
Interestingly, close to the moduli space boundary the solutions of the collective coordinate model present a self-similar nature. Of course,  as these solutions very accurately describe the true dynamics, one can draw the conclusion that the full solutions of the model have this self-similar form close to the annihilation/creation point. 

Let us start with the expansion of the BPS solutions close to the boundary i.e., where $a \to -\infty$, 
\be
\Phi_{\infty}(x;a) \approx  2\pi -\frac{2}{|x|-a}= 2\pi +\frac{2a^{-1}}{\left| \frac{x}{a} \right| +1}
\ee
One has to remember that the true BPS solution is a smooth configuration at any $x$, while this approximation is of a peakon type i.e., it possesses a spike at the origin. The deviation (in the sense of the functional norm) between the BPS solution and the approximation decreases as we approach the boundary. 

From the moduli space dynamics we know that $a(t)=- (v_bt)^{-2}$, where we put $t_b=0$. Hence
\be
\Phi_{\infty}(x,t) \approx 2\pi -\frac{2t^2}{\left| t^2 x \right|+v_b^{-2}} \equiv 2\pi - \delta \Phi_{\infty}(x,t)
\ee
where $\delta \Phi_{\infty}(x,t)$ 
is a function obeying a generalized self-similarity
\be
\delta \Phi_{\infty}(x,t) = \lambda^{-2} \delta \Phi_{\infty}(\lambda^{-2} x, \lambda t)
\ee
Often, self-similar solutions are a signature of the formation of a singularity from smooth initial data in a finite time evolution \cite{bizon}, see also \cite{rev} for a review.  One may relate such a singularity with the moduli space boundary or, equivalently, the creation/annihilation point. However, it should be underlined that the self-similar solutions never reveal any blow up behaviour and do not lead to a true singularity in the functional space. Simply, as we depart from the creation point, the solutions slowly lose their self-similar property and transform into a well separated antikink-kink pair. 

This allows us to explain why it is quite difficult to create an antikink-kink pair from generic initial data. Such a pair can only show up if the initial configuration is close to this self-similar profile. Then, it evolves toward a configuration of a separated antikink and kink which move into opposite directions. If the initial state is far from the self-similar form, then we observe a decay into radiation, that is, wave packets. 
\section{Explanation and generalizations}

Now we explain why this particular model leads to a moduli space with a boundary while for many other theories (and processes), also for many BPS-impurity models, moduli spaces are either manifolds without boundary, or the boundary cannot be attained in a finite time (or energy) evolution. 

Let us analyze the Bogomol'nyi equation (\ref{Bog-imp-sG2}) and rewrite it as
\be
\phi_x=  \tanh(x) W, \;\;\; W=2\sin^2\frac{\phi}{2},
\ee
where the pre-potential is a {\it non-negative} function taking the value zero only at the vacua $\phi=0,2\pi$. This implies that for $x<0$, $\phi$ is a monotonously decreasing function while for $x>0$ it is a monotonously increasing function. At the origin the field takes a minimal value. Only such fields solve the Bogomol'nyi equation. Since we have {\it only one} Bogomol'nyi equation, we conclude that  we can have only an antikink to the left and a kink to the right. No other configurations are possible. This is why the moduli space has a boundary which is the vacuum $\phi=2\pi$. And as the vacuum has a finite distance $s$ from any other member of the moduli space it can be approached in a finite time. 

This immediately allows us to find a MS with accessible boundary in various other models. The first class of theories with this property is any BPS-impurity model (\ref{BPS-imp}) with a non-negative $W$, for example $W=(1-\phi^2)^2$. 

Note that, if the pre-potential can change the sign, as in the case of the BPS-impurity $\phi^4$ model, where $W=1-\phi^2$, then the moduli space can be enlarged beyond the point describing the vacuum. This additional part of the moduli space is smoothly joined with the usual one and no boundary exists \cite{BPS-imp-phi4}. Physically, configurations of this new part can be related to a situation where solitons passed through each other, see \cite{BPS-imp-phi4} or \cite{MORW-2} for details. 
\\
For the BPS-impurity model discussed here, such a situation can be realized by a slightly different choice of the pre-potential
\be
W=2\sin \frac{\phi}{2}  \left| \sin \frac{\phi}{2} \right|.
\ee
This model has a different symmetry, $\phi \to \phi+4\pi$. The fundamental domain is now $[0,4\pi]$ and there is a new kink $K^*$ interpolating from $2\pi$ to $4\pi$. As a consequence, we find a new part of the moduli space describing a pair of $K^*$ and $AK^*$ at arbitrary separation. Hence, there is no boundary and the full moduli space covers the $AK+K \to K^*+AK^*$ process. 

The second class of theories, without any background field, which may admit a moduli space with an accessible boundary are BPS multi-scalar field theories. This happens provided the following conditions are fulfilled (for concreteness for a model with two scalars):  {\it (i)} There should exist a BPS sector with two zero modes. This means that the model has two (coupled) Bogomol'nyi equations, each for each field, with solutions forming a two-parameter family; {\it (ii)} Not all signs of the Bogomol'nyi equations are possible. The last requirement clearly occurs in our one-field example, see eq. (\ref{Bog-eq}) and (\ref{Bog-imp}). Instead of both signs for the usual scalar models, $\phi_x=\pm W^2$, we have only one sign in the BPS-impurity model, $\phi_x=+\sigma(x)W$. Equivalently, this condition means that the model possesses {\it chiral} Bogomol'nyi equation(s) which result in {\it chiral} solitons (kinks). While there are a lot of theories obeying the first requirement \cite{B-1, SV, AA-1, AA-2, FKZ, LT}, much less is know about models with chiral BPS kinks. However, we are aware of one such model \cite{SW-2field}. Namely, 
\bea
\mathcal{L}[\phi,\psi]&=& \frac{1}{2} \left(\partial_\mu \phi \right)^2 - \frac{1}{2} (\phi^2-1)^2 +\frac{1}{2} \left(\partial_\mu \psi \right)^2 \nonumber \\
&-& \frac{1}{2}  \phi^2(\psi^2-1)^2 + \phi (1-\psi^2)\partial_x \psi
\eea
It does have chiral Bogomol'nyi equations
\bea
\frac{d\phi}{dx} \pm (1-\phi^2) &=& 0, \nonumber \\
\frac{d\psi}{dx} - \phi(1-\psi^2) &=& 0.
\eea
Hence, replacing everywhere $(1-\psi^2)$ by a non-negative function of $\psi$, e.g., $(1-\psi^2)^2$ or $(1-\cos\psi)$, will lead to BPS solitons that form a moduli space with an accessible boundary. 

Finally, let us comment that there is a very important class of theories with a chiral Bogomol'nyi equation. These are magnetic planar Skyrme models at critical coupling, first introduced in \cite{Sch} and then further studied in \cite{SB, W, Sch-2, H}. Here the Bogomol'nyi equation can by put in a chiral form, which is a gauged version of the Bogomol'nyi equations in the standard Belavin-Polyakov model, where the gauge field is, however, a fixed background field. Therefore, there is a close relation to planar BPS soliton-impurity models \cite{BPS-imp-susy, jose}. We think that for a suitably chosen background (gauge) field one should find a moduli space with an accessible boundary. 

\section{Summary}

In the present work we have shown that the moduli space concept can be extended to cases where the moduli space has a boundary which can be attained in a finite time. This property should not be treated as a pathology which would disqualify a related collective coordinate model. As we have demonstrated, instead, it can be related to physically sound phenomena as, e.g., a complete annihilation (or, due to the time reflection symmetry, a creation) of a kink-antikink pair. This has been discussed in a particular model, but we explained that such moduli spaces may exist for rather broad classes of solitonic models, potentially also in higher dimensions. 

Interestingly, we also found that in the vicinity of the boundary of the moduli space, i.e., close to the creation/annihilation point, the solutions reveal an approximate self-similar structure. Such self-similar solutions are often associated with a singularity, which in our case is the boundary of the moduli space. The existence of these solutions explains the observed difficulties in the creation of pairs of solitons from generic initial data. 

It would be very interesting to see what happens with the boundary of the one-dimensional moduli space considered in this letter if we add more degrees of freedom, that is, if we go beyond the energetically equivalent BPS solutions to more general configurations. Then the dynamics of the resulting CCM is not just a geodesic flow on the moduli space, but important modifications can arise from a non-trivial effective potential. It would be desirable to add new degrees of freedom, which could parametrize field configurations which are not kink-confined. This could model the lumps (radiation) emitted in the annihilation, which cannot be related to any internal modes of the kinks. 

In fact, the coupling of radiative degrees of freedom to a CCM is still an open issue. We believe that the models with an accessible boundary may serve as useful laboratories to study this problem. Here, at the moment of the annihilation, the energy stored in the kink degrees of freedom is completely transferred to radiation. 

Another open problem is how quantum corrections influence the accessible boundary. Almost certainly, they will lead to the appearance of an effective potential, as usually happens for any BPS solutions \cite{AA-q, Weigel, Jarah}. Probably, the most promising way to study this question is to apply the manifestly finite, Hamiltonian one-loop approach developed in  \cite{Jarah-1,Jarah-2}. 

\vspace*{0.5cm}

\section*{Acknowledgements}

C. A. and A. W. 
acknowledge financial support from the Ministry of Education, Culture, and Sports, Spain (Grant No. PID2020-119632GB-I00), the Xunta de Galicia (Grant No. INCITE09.296.035PR and Centro singular de investigación de Galicia accreditation 2019-2022), the Spanish Consolider-Ingenio 2010 Programme CPAN (CSD2007-00042), and the European Union ERDF.

K. O., T. R., and A. W. were supported by the Polish National Science Centre (Grant No. NCN 2019/35/B/ST2/00059).

\end{document}